\documentclass[twocolumn,aps,prl,preprintnumbers]{revtex4}
\usepackage{bbm}
\usepackage{amsmath}
\usepackage{amssymb}
\usepackage{graphicx}
\usepackage{mathrsfs}
\usepackage{amsfonts}
\usepackage{amsthm}
\usepackage{color}
\usepackage{txfonts}
\usepackage[colorlinks=true,citecolor=blue,linkcolor=blue,urlcolor=blue,anchorcolor=blue]{hyperref}%
\usepackage{pifont}
\hypersetup{colorlinks=true,citecolor=blue,linkcolor=blue,urlcolor=blue}
\providecommand{\U}[1]{\protect\rule{.1in}{.1in}}
\makeatletter
\@ifundefined{textcolor}{}
{
\definecolor{BLACK}{gray}{0}
\definecolor{WHITE}{gray}{1}
\definecolor{RED}{rgb}{1,0,0}
\definecolor{GREEN}{rgb}{0,1,0}
\definecolor{BLUE}{rgb}{0,0,1}
\definecolor{CYAN}{cmyk}{1,0,0,0}
\definecolor{MAGENTA}{cmyk}{0,1,0,0}
\definecolor{YELLOW}{cmyk}{0,0,1,0}
}

\makeatother
\begin{document}
\title{Domain wall skyrmion-based magnonic crystal}

\author{Zhenyu Wang$^{1}$}
\author{Xingen Zheng$^{2}$}
\email[Corresponding author: ]{zhengxingen@sslab.org.cn}
\author{Zhixiong Li$^{3}$}
\email[Corresponding author: ]{zhixiong\_li@csu.edu.cn}
\author{Zhizhi Zhang$^{4}$}
\author{Xiansi Wang$^{1}$}

\affiliation{$^{1}$School of Physics and Electronics, Hunan University, Changsha 410082, China\\
$^{2}$Songshan Lake Materials Laboratory, Dongguan, Guangdong 523808, China\\
$^{3}$School of Physics, Central South University, Changsha 410083, China\\
$^{4}$School of Mechanical and Electrical Engineering, Chengdu University of Technology, Chengdu 610059, China}

\begin{abstract}
Magnonic waveguide based on domain wall (DW) is considered as a crucial breakthrough toward the realization of magnonic nanocircuits. However, the effective control of spin waves propagating in DWs remains to be explored. Here, we construct a magnonic crystal (MC) by using a chain of the domain wall skyrmions (DWSKs) to manipulate the spin-wave propagation in DWs. We show that the DWSK chain can be created by leveraging voltage-controlled Dzyaloshinskii-Moriya interaction. The DWSK-based MC opens magnonic bandgaps, which can be dynamically adjusted through magnetic fields modulating the DWSK size. Furthermore, the manipulation of spin waves by the DWSK-based MC maintains robust in curved DW, demonstrating its adaptability to complex device architectures. Our work provides an effective method to control the spin-wave propagation in DWs and paves the way for designing energy-efficient magnonic nanocircuits.
\end{abstract}

\maketitle
\section{Introduction}\label{sec1}
Magnons, the quanta of spin waves, are considered as potential data carriers for future information procession and logic operation, due to their advantages of no Joule heating, short wavelengths down to nanometer scale, and a wide frequency range spanning from GHz to THz \cite{Serga2010,Chumak2015,Barman2021,Chumak2022}. Controlled propagation and manipulation of spin waves at the nanoscale is crucial for the miniaturization and integration of future magnonic devices.
Magnetic domain walls (DWs), which are boundaries between different magnetic domains, can be as narrow as tens of nanometers in width \cite{Lloyd2001, Moreno2016}. One-dimensional potential well caused by the DW can serve as an internal channel to guide the spin-wave propagation \cite{Winter1961,Lan2015,Xing2016}, which has already been predicted theoretically \cite{Garcia2015,Hartmann2021,Liang2022} and demonstrated experimentally \cite{Wagner2016,Albisetti2018,Sluka2019}. Compared to magnetic strip-based magnonic waveguides,
DW-based magnonic waveguides exhibit remarkable advantages, such as narrow spin-wave beams \cite{Xing2016,Wagner2016}, reduced energy loss \cite{Hartmann2021}, adaptation to curved geometry \cite{Garcia2015,Albisetti2018}, faster group velocities \cite{Garcia2015,Park2021}, and ease of reconfiguration \cite{Wagner2016,Albisetti2018}.
Despite the unique advantages of DWs in guiding spin waves, the manipulation of spin-wave propagation within DWs remains largely unexplored.

Magnonic crystal (MC) provides an effective method to manipulate spin-wave propagation \cite{Krawczyk2014,Chumak2017,Zakeri2020,Lv2025}. It can significantly alter the magnonic band structure, resulting in the formation of allowed passbands and forbidden bandgaps, effectively functioning as a spin-wave filter \cite{Kim2009,Merbouche2021}. The original MCs are constructed through a periodic modulation of material parameters \cite{Wang2009,Ciubotaru2013,Li2016,Qin2018,Mieszczak2022,Adhikari2023} or structural geometries \cite{Lee2009,Frey2020,Bir2024,Mantion2024,Sadovnikov2022,Korniienko2019}.
However, these material- or geometry-modulated MCs are hard to tune, once they are fabricated.
Several methods have been proposed to realize the reconfigurable MCs, including a spatially periodic magnetic field produced by a current-carrying planar metallic meander structure \cite{Chumak2009,Karenowska2012}, spatial modulation of saturation magnetization through optically induced thermal landscapes \cite{Vogel2015}, voltage-controlled magnetic anisotropy \cite{Wang2017}, and a periodic strain induced by a surface acoustic wave \cite{Chumak2010}. Besides, magnetic soliton (e.g. domain wall \cite{Li2015,Wang2015}, skyrmion \cite{Ma2015,Chen2021}, and hopfion \cite{Medlej2024}) based MCs have recently attracted much attention due to their dynamically tunability \cite{Li2015,Ma2015,Wang2020} and nontrivial topological characteristics \cite{Li2018,Li202102,Diaz2019,Hirosawa2020,Xie2021}.

Most recently, a type of MC has been proposed to control spin waves in DWs, constructed by dipolar-coupled ferromagnetic bilayer structures \cite{Yi2024}. However, adding a second ferromagnetic layer complicates fabrication and hinders the integration of magnonic devices. In fact, DWs also possess finer topological substructures, including Bloch lines \cite{Malozemoff1972,Yoshimura2016}, domain wall bimerons \cite{Nagase2021,Chen2025,Fu2025}, and domain wall skyrmions (DWSKs) \cite{Cheng2019,Li2021,Yang2021,Ross2023,Amari2024}. The DWSK is topologically equivalent to the conventional skyrmion and can be considered as a skyrmion trapped inside the DW. Recent studies have demonstrated that the DWSK driven by the current moves along the DW, with its skyrmion Hall effect being mitigated by the DW confinement \cite{Han2024,Nie2025,Xiao2025}. Due to this advantage, the DWSK is regarded as a superior information carrier in racetrack memories compared to the skyrmion. Moreover, recent studies on magnonics in DWs containing bimeron chains have uncovered topologically protected magnonic edge states \cite{Saji2025}.
Thus, DWs hosting finer topological objects offer a novel platform for exploring intriguing phenomena and manipulating spin-wave propagation.

In this work, we construct a MC by using a series of periodically distributed DWSKs (hereafter referred to as ``DWSK-MC"), which can be created by locally applying a voltage pulse. Then, the magnonic spectrum in the DWSK-MC is investigated both theoretically and numerically, showing that the bandgap of spin waves propagating in the DW is opened. Such a manipulation of spin waves using the DWSK-MC is effective even in curved DWs. Furthermore, the DWSK size can be dynamically adjusted through a bias magnetic field, which modifies the periodic potential well. This highlights the dynamic tunability of the proposed DWSK-MC.

\section{Spin configuration of the DWSK-MC}

\begin{figure}
    \centering
    \includegraphics[width=1\linewidth]{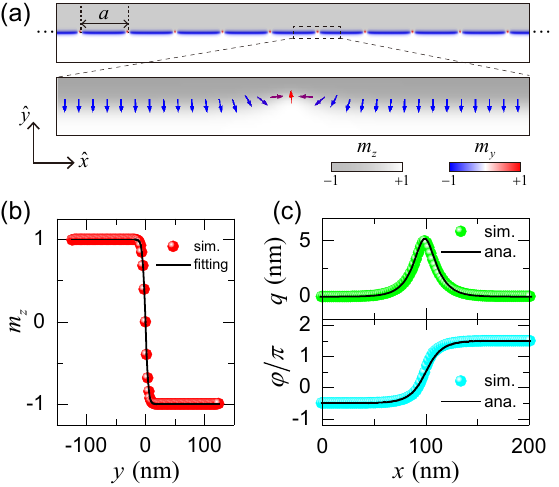}
    \caption{(a) Schematic image of a DWSK-MC. The distance between two nearest-neighbor DWSKs is $a$. The inset in (a) is the enlarged image of the magnetization vector of the dashed rectangular region. (b) The $z$-component magnetization along the $\hat{y}$ direction. The red dots are simulation results and the black curve is the fitting line based on the Walker profile.
    (c) The deformation and azimuthal angle of the DWSK in the inset of (a). The solid black curves are plotted based on Eq. (\ref{eq_DWSK_profile}) and the color symbols are obtained from micromagnetic simulations.}
    \label{fig1}
\end{figure}

We consider a chiral magnetic strip, with its energy density expressed as
\begin{equation}\label{eq_Etot}
     \varepsilon_{\mathrm{tot}}=A(\nabla\mathbf{m})^{2}-D\mathbf{m}\cdot[(\hat{z}\times\nabla)\times\mathbf{m}] -K m_{z}^{2},
\end{equation}
where $\mathbf{m}=\mathbf{M}/M_s$ is the normalized magnetization with the saturation magnetization $M_s$, $A$ is the exchange constant, $D$ is the magnitude of Dzyaloshinskii-Moriya interaction (DMI) with the interfacial form, $K=K_u-\mu_0 M_s^2/2$ is the effective perpendicular magnetic anisotropy with $K_u$ the uniaxial anisotropy along the $\hat{z}$ axis and the vacuum permeability $\mu_0$. A N\'{e}el-type DW, hosting a series of periodically distributed DWSKs, is located in the center of the magnetic strip, as shown in Fig. \ref{fig1}(a). We express the unit magnetization in spherical coordinates as $\mathbf{m}=\{\sin\theta\cos\varphi, \sin\theta\sin\varphi, \cos\theta\}$. Then, the magnetization profile of the DWSK can be described by the modified Slonczewski ansatz \cite{Cheng2019,Han2024,Nie2025}:
\begin{eqnarray}\label{eq_DWSK_profile}
\begin{aligned}
  \theta(x,y)&=2\arctan\Big[\exp\Big(\frac{y-q(x)}{\Delta}\Big)\Big],\\
  \varphi(x) &=4\arctan\Big[\exp\Big(\frac{x-x_s}{\Delta_s}\Big)\Big]-\frac{\pi}{2}, \\
  q(x) &= 2\kappa\Delta_s \mathrm{sech}\Big(\frac{x-x_s}{\Delta_s}\Big),
\end{aligned}
\end{eqnarray}
where $q(x)$ is the DW position deviating from $y=0$ and $\varphi(x)$ is the azimuthal angle of the magnetization inside the DW. $x_s$ and $\Delta_s=\Delta/\sqrt{\kappa}$ are respectively the position and size of the DWSKs with $\Delta=\sqrt{A/K}$ and $\kappa=\pi D/(4\sqrt{A K})$.

To verify the spin configuration of DWSKs, micromagnetic simulations are performed using Mumax3 \cite{Vansteenkiste2014} with the magnetic parameters of Co: $M_s=5.8\times10^{5}$ $\mathrm{A/m}$, $A=15$ $\mathrm{pJ/m}$, $K_{u}=8\times10^{5}$ $\mathrm{J/m^3}$, and $D=1$ $\mathrm{mJ/m^2}$. The damping is set as $\alpha=0.1$ in simulations for obtaining the equilibrium static magnetization and creating the DWSK-MC. For the spin-wave propagation in the DWSK-MC, the damping is set as a smaller value $\alpha=0.01$ to facilitate a long-distance propagation.
A magnetic strip with the length 4000 nm, width 250 nm, and thickness 2 nm is discreted by the cell size of $2\times2\times2$ $\mathrm{nm}^3$. At the strip center, a DWSK-MC with the periodic distance $a$ is formed, as shown in Fig. \ref{fig1}(a). The enlarged image of a DWSK is plotted in the inset of Fig. \ref{fig1}(a), showing a slight deformation of the DW profile. By fitting the $z$-component magnetization ($m_z$) with the Walker profile [see Fig. \ref{fig1}(b)], we can obtain the DW position and extract the azimuthal angle of the DW magnetization, as plotted in Fig. \ref{fig1}(c). One can see that simulation results agree excellently with the analytical formula Eqs. (\ref{eq_DWSK_profile}), thereby validating the magnetization profile of the DWSK.

\section{The DWSK-MC creation}

To generate the DWSK-MC, the DW magnetization is required to be locally reversed. The magnetization orientation of the chiral DW is determined by the DMI, whose sign can be reversed by a voltage pulse \cite{Ma2024}. Recent studies have demonstrated that the voltage-controlled DMI torque can effectively switch both perpendicular and in-plane magnetizations \cite{Yu2023,Yu2024}. Thus, we utilize the voltage-controlled DMI to create the DWSK-MC.

\begin{figure}
    \centering
    \includegraphics[width=1\linewidth]{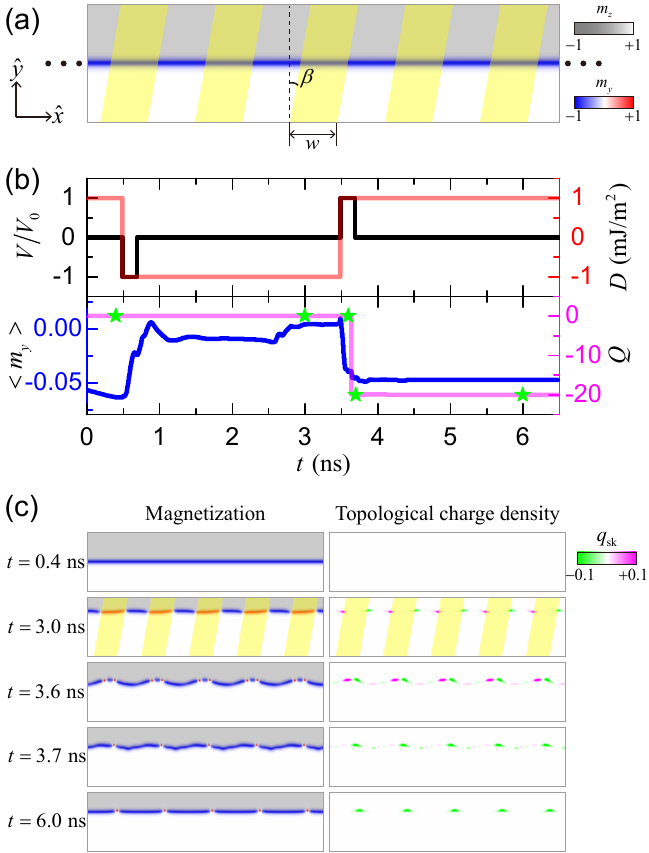}
    \caption{(a) Top view of a N\'{e}el-type DW located in the strip integrated with an array of the electrodes for applying the voltage pulse. (b) Upper panel: Time evolution of the voltage pulse (black line) and DMI constant (red line). Bottom panel: The averaged $y$-component magnetization over the whole strip $\langle m_y\rangle$ (blue line) and topological charge $Q$ (magenta line) as a function of $t$. (c) The spatial distributions of the magnetization and topological charge density ($q_{\mathrm{sk}}$) at selected times labeled as green stars in (b). The yellow regions in (a) and (c) represent the electrode positions.}
    \label{fig2}
\end{figure}

Figure \ref{fig2}(a) shows a N\'{e}el-type DW located in the strip center. Then, a series of periodically distributed local voltage pulse is applied on the strip. The width of the electrode is $w=100$ nm and the separation between two nearest electrodes is 200 nm. $\beta$ is the rotating angle of the electrodes with respect to the $y$ axis.
For $\beta=0$, the nucleation of the DWSK-MC is uncontrollable due to the random deformation of the DW (see Appendix \ref{Appendix_A}).
Fortunately, the DW deformation can be adjusted by rotating the electrodes. By tuning the rotating angle of the electrode, a consistent deformation of the DW can be achieved, thereby making the creation of the DWSK-MC controllable.

Here, we take the electrode with $\beta=-10^{\circ}$ as an example to illustrate the nucleation of the DWSK-MC. When a voltage pulse is switched on at 0.5 ns, the DMI sign is reversed, as shown in the upper panel of Fig. \ref{fig2}(b). Then, the magnetization below each electrode switches from $m_y=-1$ to $m_y=+1$ [see the bottom panel in Fig. \ref{fig2}(b) and snapshot at 3.0 ns in Fig. \ref{fig2}(c)]. At 3.5 ns, an opposite voltage pulse is applied, the DMI sign reverses back. The switched magnetization ($m_y=+1$) below each electrode transforms to two DWSKs with opposite topological charges $Q=\pm1$ at 3.6 ns. It is known that the DWSKs with $Q=\pm1$ correspond to the DW down-bending and up-bending, respectively \cite{Nie2025}. Therefore, the up-bent DW prefers the DWSKs with $Q=-1$. The DWSKs with $Q=+1$ are unstable and annihilated. Finally, a chain of DWSKs with $Q=-1$ is nucleated in the DW at 6.0 ns.

The topological charge is also calculated by $Q=(1/4\pi)\iint q_{\mathrm{sk}}dxdy$ with the topological charge density $q_{\mathrm{sk}}=\mathbf{m}\cdot(\partial_x\mathbf{m}\times\partial_y\mathbf{m})$, as depicted in the bottom panel of Fig. \ref{fig2}(b). One DWSK with $Q=-1$ can be generated under each electrode.
Thus, the electrodes spaced at 200 nm intervals can produce 20 DWSKs along a 4000 nm length strip, resulting in a topological charge of $Q=-20$ in Fig. \ref{fig2}(b).
Furthermore, reversing the rotating angle of the electrodes ($\beta=10^{\circ}$) generates a chain of DWSKs with
an opposite topological charge $Q=+1$ (see Appendix \ref{Appendix_A}). These results indicate that the topological charges of the DWSK-MC can be controlled by adjusting the electrode rotation. The specific relationship between the DWSK creation and rotating angles is beyond the scope of this work and will be addressed in future studies. Additionally, other methods for generating the DWSK, such as magnetic field, spin-polarized current, strain, and laser pulses, merit further investigation.

\section{Spin-wave propagation in the DWSK-MC}

\begin{figure}
    \centering
    \includegraphics[width=1\linewidth]{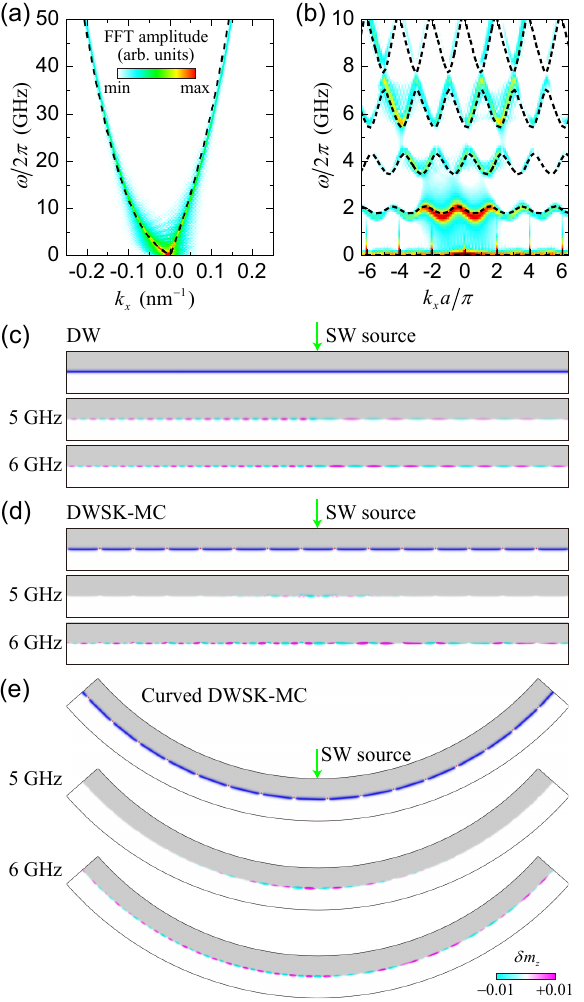}
    \caption{The dispersion relation of spin waves propagating in (a) DW and (b) DWSK-MC with the periodic length $a=200$ nm. The intensity of spin waves is indicated by the color scale. The black dashed curves are the analytical results. (c-e) Spatial profiles of spin waves at 5 GHz and 6 GHz, propagating in (c) DW, (d) straight DWSK-MC, and (e) curved DWSK-MC. The green arrows in (c)-(e) are the exciting sources of spin waves.}
    \label{fig3}
\end{figure}

In the following, we proceed to study the spin-wave propagation in the DWSK-MC. To obtain the spin-wave dispersion relation, a sinc-function field $\mathbf{h}(t)=h_{0}\mathrm{sinc}[\omega_c(t-t_0)]\hat{z}$ with amplitude $h_0=10$ mT and $t_0=1$ ns is locally applied over a 10 nm wide rectangular region in the middle of the strip. The cutoff frequency is $\omega_c/2\pi=50$ GHz, which is below the frequency gap ($\gamma K/\pi M_s=56.8$ GHz) of bulk spin waves outside the DW. The band structure of bound spin waves can be obtained by the fast Fourier transform (FFT) of the spatiotemporal oscillation of the $z$-component magnetization ($\delta m_z$) within the DW.
As reference, we plot the dispersion relation of spin waves propagating in a pure DW without the DWSKs [see Fig. \ref{fig3}(a)]. One can see that bound spin waves in the N\'{e}el-type DW are gapless and nonreciprocal, which can be well described by the analytical formula \cite{Garcia2015}
\begin{equation}\label{eq_DWSW_dispersion}
  \omega=\sqrt{\omega_k(\omega_k-\omega_{\perp}+\frac{\omega_{D}}{k_{x}\Delta})}-\omega_{D},
\end{equation}
where $\omega_k=2\gamma A k_x^2/M_s$, $\omega_{D}=\pi\gamma D k_x/2 M_s$, $\omega_{\perp}=2\gamma K_{\perp}/M_s$, $K_{\perp}=\mu_0 N_y M_s^2/2$, and $N_y=L_z/(L_z+\pi\Delta)$ with $L_z$ being the film thickness.

In the DWSK-MC with a periodic length $a=200$ nm, the DWSKs generate a periodic potential well, which would open the magnonic bandgaps in DW. As expected, the magnonic band in the DWSK-MC changes significantly, where the forbidden bands of spin waves can be clearly observed [see Fig. \ref{fig3}(b)]. Furthermore, it is noted that bandgaps are not opened at the Brillouin zone (BZ) boundaries ($k_x=n\pi/a$), which is due to the nonreciprocal propagation of spin waves in chiral DWs \cite{Garcia2015}. Different wavelengths of two counterpropagating spin waves shift the minimum and maximum of the dispersion branches out of the borders of the BZs, giving rise to the indirect bandgaps, which has also been demonstrated in Refs. \cite{Gallardo2019,Wei2025}. The magnonic band in the DWSK-MC is also theoretically calculated by using the Holstein-Primakoff transformation and diagonalizing the magnon Hamiltonian through the para-unitary transformation (see Appendix \ref{Appendix_B} for detailed derivations), showing a good agreement with the simulation results.
Notedly, a minor deviation between the analytical and simulation results is observed, potentially due to the magnon-driven DWSK motion resulting from the momentum exchange \cite{Xiao2025}.

To examine the manipulation of spin waves by the DWSK-MC, we apply a sinusoidal monochromatic microwave source $\mathbf{H}_{\mathrm{ext}}=h_{0}\sin(\omega t)\hat{z}$ at the middle of the strip. We choose two representative frequencies $\omega/2\pi=5$ GHz and 6 GHz, which are located in the forbidden and allowed bands, respectively. In the pure DW, spin waves with both frequencies can propagate but with different wavelengths along $\pm x$ directions, which is caused by the DMI, as shown in Fig. \ref{fig3}(c). For the DWSK-MC, spin waves with 5 GHz are prohibited from propagation, while it permits the propagation of spin waves with 6 GHz [see Fig. \ref{fig3}(d)]. This result demonstrates the manipulation of the spin-wave propagation in DWs through the DWSK-MC.
To verify the robustness of the DWSK-MC, we bend the strip into a curved geometry with an arc angle of approximately equal to $100^{\circ}$ and a radius of curvature of 2000 nm for the outer edge. It is observed that the spin-wave propagation in curved DWs is similar to the case of the straight DWs [see Fig. \ref{fig3}(e)], which suggests that the manipulation of spin waves by the DWSK-MC remains feasible in curved DWs.

\section{Field-controlled DWSK-MC}

\begin{figure}
    \centering
    \includegraphics[width=1\linewidth]{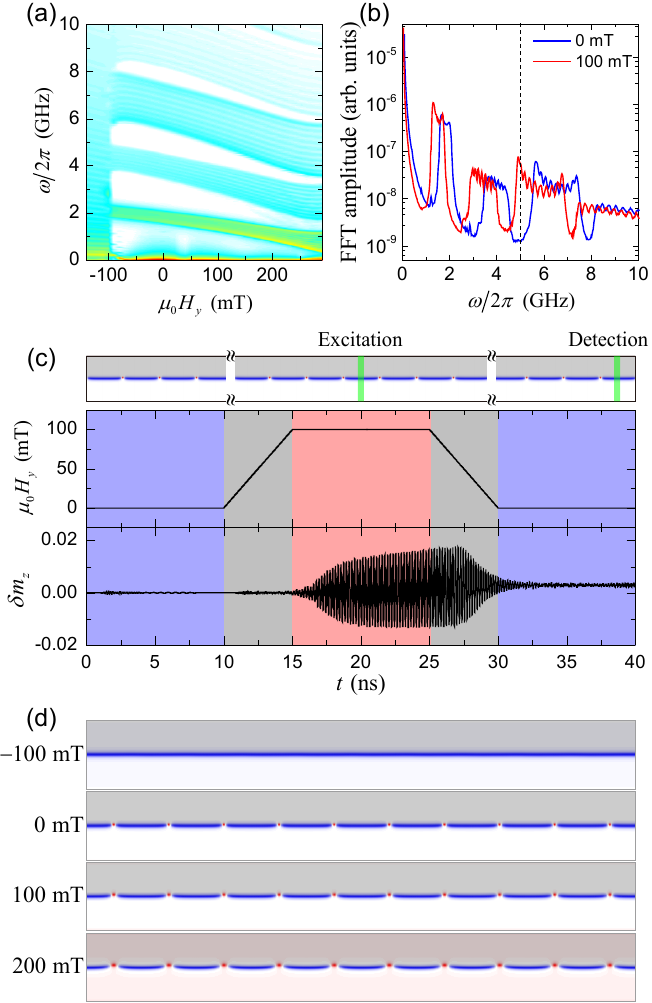}
    \caption{(a) The magnonic band structure as a function of the applied magnetic field ($\mu_{0} H_{y}$) along the $\hat{y}$ direction. (b) The FFT spectrum of spin waves in DWSK-MC under $\mu_{0} H_{y}=0$ and 100 mT. The dashed black line labels the frequency of 5 GHz. (c) The time evolution of the external magnetic field $\mu_{0} H_{y}$ and the magnetization oscillation $\delta m_{z}$ at the detection position. (d) Magnetization states of the strip under different magnetic field strengths.}
    \label{fig4}
\end{figure}

The band structure of spin waves propagating in the DWSK-MC can be dynamically tuned by magnetic field along the $\hat{y}$ direction, as shown in Fig. \ref{fig4}(a). Upon decreasing the magnetic field from 290 mT to -90 mT, the widths and center frequencies of bandgaps increase. Such a field dependence of the magnonic band structure is due to the variation of the DWSK size, which increases with the field strength [see Fig. \ref{fig4}(d)]. It would affect the periodic potential well felt by spin waves, leading to the modification of the magnonic band structure. When the magnetic field is lower than -100 mT, the DWSKs are annihilated and the bandgaps disappear, as shown in Figs. \ref{fig4}(a) and \ref{fig4}(d). Figure \ref{fig4}(b) shows the spectra of spin waves propagating in DWSK-MC under two different field strengthes. One can see that the external magnetic field shifts the magnonic band structure to the low frequency regime. Under two different field strengthes, some magnon modes with the same frequency are located in forbidden and allowed bands, respectively. For instance, spin wave with 5 GHz [labeled by the dashed black line in Fig. \ref{fig4}(b)] resides within the bandgaps for 0 mT and shifts into the allowed bands for 100 mT.

To further illustrate the dynamical modulation of the DWSK-MC by the external field, we examine the spin-wave propagation during the process of the applied external field changing from 0 mT to 100 mT, and then returning to 0 mT. To avoid the drastic magnetization oscillation induced by the abrupt change of the external field, the field amplitude is linearly varied by 100 mT in 5 ns [see the gray region of the lower panel in Fig. \ref{fig4}(c)]. Spin wave with 5 GHz is excited at the middle of the strip and detected at the right side of the strip 1600 nm away from the center [see the green bars of the upper panel in Fig. \ref{fig4}(c)].
The detected signal is plotted in the lower panel of Fig. \ref{fig4}(c). At $0\sim10$ ns, the external field is turned off, spin wave with 5 GHz is located in bandgaps and cannot propagate in the DWSK-MC. Thus, no signal is detected at this stage. When the field amplitude increases to 100 mT ($15\sim25$ ns), spin wave with 5 GHz is transferred to the allowed bands and can propagate in the DWSK-MC, which is verified by the significant detected signal with the coherent oscillation [see the red region of the lower panel in Fig. \ref{fig4}(c)]. When the field amplitude decreases back to 0 mT ($30\sim40$ ns), spin wave with 5 GHz returns to the bandgaps, resulting in the disappearance of the signal at the detection. This result suggests that the spin-wave propagation in the DWSK-MC can be dynamically tuned by the external field.

\section{Conclusion}
In summary, we have demonstrated a dynamic MC based on a periodically distributed DWSKs inside the DW-based magnonic waveguide. The DWSK-MC can be created by applying a local voltage pulse through the voltage-controlled DMI effect. The periodic modulation of the potential well induced by the DWSKs leads to a pronounced modification of the magnonic band, resulting in the opening of the magnonic bandgaps. Moreover, the DWSK-MC functions well in curved DWs, which is more feasible for applications in magnonic circuits with complex shapes. Another advantage of the DWSK-MC is its dynamic controllability. The width and central frequencies of the magnonic bandgaps can be tuned by changing the DWSK size via applying the external magnetic field. It is worth mentioning that the MC proposed in this work can also be constructed by using other DW solitons, including Bloch lines and DW bimerons. Our finding offers an efficient approach for manipulating spin waves in DWs, which would promote the development of magnonic nanocircuits.

\section*{ACKNOWLEDGMENT}
We thank R. Wang and H. Yuan for helpful discussions. This work is supported by the Fundamental Research Funds for the Central Universities. Z.W. acknowledges the support the Natural Science Foundation of China (NSFC) (Grant No. 12204089) and the Natural Science Foundation of Hunan Province of China (Grant No. 2024JJ6113).
X.Z. acknowledges the support of the Guangdong Basic and Applied Basic Research Foundation (Grant No. 2024A1515110196).
Z.-X.L. acknowledges financial support from the Natural Science Foundation of Hunan Province of China (Grant No. 2023JJ40694).
Z.Z. acknowledges support from the NSFC (Grant No. 12404122), and in part by Sichuan Science and Technology Program (Grant No. 2025ZNSFSC0868).

\appendix

\section{THE DWSK CREATION VIA THE VOLTAGE-CONTROLLED DMI}\label{Appendix_A}

\begin{figure}
  \centering
  \includegraphics[width=1\linewidth]{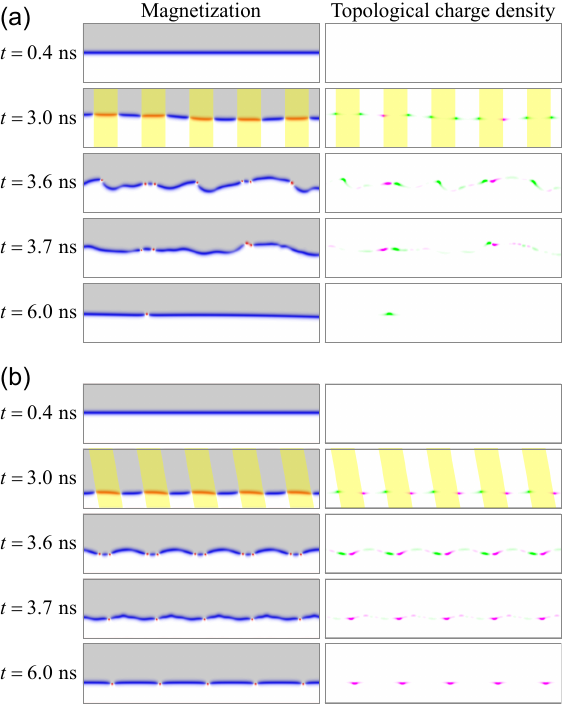}\\
  \caption{The spatial distributions of the magnetization and topological charge density during the processes of the DWSK creation for (a) $\beta=0^{\circ}$ and (b) $\beta=10^{\circ}$.}\label{figA1}
\end{figure}

Here, we present the processes involved in creating the DWSK under the rotating angles of $\beta=0^{\circ}$ and $10^{\circ}$. The local voltage pulse applied is same as that depicted in Fig. \ref{fig2}(b) of the main text.
For $\beta=0^{\circ}$, the number of the generated DWSKs is uncontrolled due to the random deformation of the DW [see Fig. \ref{figA1}(a)]. For $\beta=10^{\circ}$, one DWSK with $Q=+1$ can be generated under each electrode, as shown in Fig. \ref{figA1}(b).

\section{CALCULATION OF MAGNONIC BANDS IN THE DWSK-MC}\label{Appendix_B}
We calculate the magnonic band structure in the DWSK-MC using the Holstein-Primakoff (HP) transformation \cite{Holstein1940} and para-unitary transformation \cite{Colpa1978}, which have been widely employed in the studies of topological magnons \cite{Roldan2016,Diaz2019,Takeda2024}. We begin with a discrete spin Hamiltonian
\begin{equation}\label{B1}
  \begin{split}
  \mathcal{H}&=-J\sum_{<i,j>}\mathbf{S}_{i}\cdot\mathbf{S}_{j}-D\sum_{<i,j>}(\hat{z}\times\mathbf{\hat{r}}_{ij})\cdot(\mathbf{S}_i\times\mathbf{S}_j)
  -K\sum_i(\mathbf{S}_i\cdot\hat{z})^2 \\
  &+\frac{\mu_0\mu_s^2}{4\pi}\sum_{i<j}\frac{\mathbf{S}_{i}\cdot\mathbf{S}_{j}-3(\mathbf{S}_{i}\cdot\mathbf{r}_{ij})(\mathbf{S}_{j}\cdot\mathbf{r}_{ij})}
  {|\mathbf{r}_{ij}|^3},
\end{split}
\end{equation}
where $\mathbf{S}_i$ denotes the spin operator at the lattice site $i$, the summation is carried over all coupling spins; $\mathbf{\hat{r}}_{ij}=\mathbf{r}_{ij}/|\mathbf{r}_{ij}|$ is the unit vector pointing from site $i$ to $j$; the parameters $J$, $D$, and $K$ are the exchange constant, DMI strength, and perpendicular anisotropy constant, respectively; $\mu_0$ is the vacuum permeability and $\mu_s$ is the magnetic moment associate with each spin. These parameters are related to those in the simulation through the following relationships: $J=2A_{\mathrm{ex}}^{\mathrm{sim}}a_0/S^2$, $K=K_{u}^{\mathrm{sim}}a_0^3/S^2$, $D=D^{\mathrm{sim}}a_0^2/S^2$, $\mu_s=M_s a_0^3/S$, where $a_0=2$ nm is the lattice constant, and $S=M_s a_0^3/(g\mu_B)$ is the spin quantum number with the Bohr magneton $\mu_B$ and the Land\'{e} factor $g=2$.

Then, we consider small spin fluctuations against the background of spin textures. It is convenient to introduce a local coordinate system in which the local $z$-axis coincides with the equilibrium direction of the spin magnetization. In the local coordinate, the spin magnetization can be written as
\begin{equation}\label{B2}
  \begin{split}
     S_i^x &= \tilde{S}_i^x\cos\theta\cos\phi-\tilde{S}_i^y\sin\phi+\tilde{S}_i^z\sin\theta\cos\phi, \\
     S_i^y &= \tilde{S}_i^x\cos\theta\sin\phi+\tilde{S}_i^y\cos\phi+\tilde{S}_i^z\sin\theta\sin\phi, \\
     S_i^z &= -\tilde{S}_i^x\sin\theta+\tilde{S}_i^z\cos\theta.
   \end{split}
\end{equation}
By the HP transformation, the spin operators are expressed as
\begin{equation}\label{B3}
  \begin{split}
     \tilde{S}_i^x &= \frac{\sqrt{2S}}{2}(a_i+a_i^+), \\
     \tilde{S}_i^y &= \frac{\sqrt{2S}}{2i}(a_i-a_i^+), \\
     \tilde{S}_i^z &= S-a_i^+a_i,
   \end{split}
\end{equation}
where $a_i$ and $a_i^+$ are the annihilation and creation operators of magnons at the lattice site $i$.
Substituting Eqs. (\ref{B2}) and (\ref{B3}) into Eq. (\ref{B1}) and keeping the quadratic terms of boson operators, we have
\begin{equation}\label{B4}
  \mathcal{H}=E_0+\frac{1}{2}\sum_{ij}(A_{ij}a_i^+a_j+B_{ij}a_i^+a_j^++B_{ij}^*a_ia_j+A_{ij}^*a_ia_j^+),
\end{equation}
where $E_0$ is the ground-state energy constant, $A_{ij}$ and $B_{ij}$ denote the coupling coefficients.

Considering the periodicity of the system, we label the spins in the lattice by a combined index ($\mathbf{R},j$), where $\mathbf{R}$ indexes the unit cell and $j$ labels the site within that cell. A unit cell contains $N$ spins.
We then perform a Fourier transform of the spin operators across the crystal: $a_{\mathbf{R},j}=\sum_{\mathbf{k}}e^{i\mathbf{R}\cdot\mathbf{k}}b_{\mathbf{k},j}$, the sum runs over all wavevectors in the first Brillouin zone. Inserting this into Eq. (\ref{B4}) yields the Hamiltonian in the momentum space. After some algebra, the Hamiltonian can be written in a compact matrix form as
\begin{equation}\label{B5}
  \mathcal{H}(\mathbf{k})=\frac{1}{2}
  \left(
    \begin{array}{cc}
      \mathbf{b}_{\mathbf{k}}^{+} & \mathbf{b}_{\mathbf{-k}} \\
    \end{array}
  \right)
  \left(
    \begin{array}{cc}
      A(\mathbf{k}) & B(\mathbf{k}) \\
      B^{*}(\mathbf{-k}) & A^{*}(\mathbf{-k}) \\
    \end{array}
  \right)
  \left(
    \begin{array}{c}
      \mathbf{b}_{\mathbf{k}} \\
      \mathbf{b}_{\mathbf{-k}}^{+} \\
    \end{array}
  \right),
\end{equation}
where $\mathbf{b}_{\mathbf{k}}^{+}\equiv[b_{\mathbf{k},1},\cdots,b_{\mathbf{k},N}]$ denotes the vector of momentum-space magnon creation operators, the $N\times N$ matrices $A(\mathbf{k})$ and $B(\mathbf{k})$ describe the momentum-dependent coefficients for $\mathbf{b}_{\mathbf{k}}^{+}\mathbf{b}_{\mathbf{k}}$ and $\mathbf{b}_{\mathbf{k}}^{+}\mathbf{b}_{\mathbf{-k}}^{+}$, respectively.

The quadratic Hamiltonian in Eq. (\ref{B5}) describes a bosonic Bogoliubov-de Gennes problem for magnons, which can be diagonalized by using a para-unitary transformation. We extract the $2N\times2N$ Hermitian matrix
\begin{equation}\label{B6}
  H_{B}(\mathbf{k})=
  \left(
    \begin{array}{cc}
      A(\mathbf{k}) & B(\mathbf{k}) \\
      B^{*}(\mathbf{-k}) & A^{*}(\mathbf{-k}) \\
    \end{array}
  \right),
\end{equation}
and introduce a metric matrix $\sum_{z}=\sigma_z\otimes I_N$, where $\sigma_z$ is the third Pauli matrix and $I_N$ is the $N\times N$ unit matrix. By solving the eigenvalue problem of the pseudo-Hermitian matrix $\sum_{z}H_B$, we can obtain the eigenvalue $E(\mathbf{k})=\hbar\omega(\mathbf{k})$, i.e., magnonic bands.

In practical calculations, including all pairwise dipolar couplings is both computationally prohibitive and unnecessary due to the weakness of distant dipolar interactions. We truncate the dipolar sums beyond the eight-nearest neighbors, which yields a good approximation for the magnon spectrum.


\begin{thebibliography}{99}
\bibitem{Serga2010} A. A. Serga, A. V. Chumak, and B. Hillebrands, YIG magnonics, \href{https://doi.org/10.1088/0022-3727/43/26/264002}{J. Phys. D: Appl. Phys. \textbf{43}, 264002 (2010)}.
\bibitem{Chumak2015} A. V. Chumak, V. I. Vasyuchka, A. A. Serga, and B. Hillebrands, Magnon spintronics, \href{https://doi.org/10.1038/nphys3347}{Nat. Phys. \textbf{11}, 453 (2015)}.
\bibitem{Barman2021} A. Barman, G. Gubbiotti, S. Ladak, et. al., The 2021 Magnonics Roadmap, \href{https://doi.org/10.1088/1361-648X/abec1a}{J. Phys.: Condens. Matter \textbf{33}, 413001 (2021)}.
\bibitem{Chumak2022} A. V. Chumak, P. Kabos, M. Wu, et. al., Advances in Magnetics Roadmap on Spin-Wave Computing, \href{https://doi.org/10.1109/TMAG.2022.3149664}{IEEE Trans. Magn. \textbf{58}, 0800172 (2022)}.
\bibitem{Lloyd2001} S. J. Lloyd, N. D. Mathur, J. C. Loudon, and P. A. Midgley, Magnetic domain-wall width in $\mathrm{La_{0.7}Ca_{0.3}MnO_{3}}$ thin films measured using Fresnel imaging, \href{https://doi.org/10.1103/PhysRevB.64.172407}{Phys. Rev. B \textbf{64}, 172407 (2001)}.
\bibitem{Moreno2016} R. Moreno, R. F. L. Evans, S. Khmelevskyi, M. C. Mu\~{n}oz, R. W. Chantrell, and O. Chubykalo-Fesenko, Temperature-dependent exchange stiffness and domain wall width in Co, \href{https://doi.org/10.1103/PhysRevB.94.104433}{Phys. Rev. B \textbf{94}, 104433 (2016)}.
\bibitem{Winter1961} J. M. Winter, Bloch Wall Excitation. Application to Nuclear Resonance in a Bloch Wall, \href{https://doi.org/10.1103/PhysRev.124.452}{Phys. Rev. \textbf{124}, 452 (1961)}.
\bibitem{Lan2015} J. Lan, W. Yu, R. Wu, and J. Xiao, Spin-Wave Diode, \href{https://doi.org/10.1103/PhysRevX.5.041049}{Phys. Rev. X \textbf{5}, 041049 (2015)}.
\bibitem{Xing2016} X. Xing and Y. Zhou, Fiber optics for spin waves, \href{https://doi.org/10.1038/am.2016.25}{NPG Asia Mater. \textbf{8}, e246 (2016)}.
\bibitem{Garcia2015} F. Garcia-Sanchez, P. Borys, R. Soucaille, J. P. Adam, R. L. Stamps, and J. V. Kim, Narrow Magnonic Waveguides Based on Domain Walls, \href{https://doi.org/10.1103/PhysRevLett.114.247206}{Phys. Rev. Lett. \textbf{114}, 247206 (2015)}.
\bibitem{Hartmann2021} D. M. F. Hartmann, A. R\"{u}ckriegel, and R. A. Duine, Nonlocal magnon transport in a magnetic domain wall waveguide, \href{https://doi.org/10.1103/PhysRevB.104.064434}{Phys. Rev. B \textbf{104}, 064434 (2021)}.
\bibitem{Liang2022} X. Liang, Z. Wang, P. Yan, and Y. Zhou, Nonreciprocal spin waves in ferrimagnetic domain-wall channels, \href{https://doi.org/10.1103/PhysRevB.106.224413}{Phys. Rev. B \textbf{106}, 224413 (2022)}.
\bibitem{Wagner2016} K. Wagner, A. K\'{a}kay, K. Schultheiss, A. Henschke, T. Sebastian, and H. Schultheiss, Magnetic domain walls as reconfigurable spin-wave nanochannels, \href{https://doi.org/10.1038/nnano.2015.339}{Nat. Nanotechnol. \textbf{11}, 432 (2016)}.
\bibitem{Albisetti2018} E. Albisetti, D. Petti, G. Sala, R. Silvani, S. Tacchi, S. Finizio, S. Wintz, A. Cal\`{o}, X. Zheng, J. Raabe, E. Riedo, and R. Bertacco, Nanoscale spin-wave circuits based on engineered reconfigurable spin-textures, \href{https://doi.org/10.1038/s42005-018-0056-x}{Commun. Phys. \textbf{1}, 56 (2018)}.
\bibitem{Sluka2019} V. Sluka, T. Schneider, R. A. Gallardo, A. K\'{a}kay, M. Weigand, T. Warnatz, R. Mattheis, A. Rold\'{a}n-Molina, P. Landeros, V. Tiberkevich, A. Slavin, G. Sch\"{u}tz, A. Erbe, A. Deac, J. Lindner, J. Raabe, J. Fassbender, and S. Wintz, Emission and propagation of 1D and 2D spin waves with nanoscale wavelengths in anisotropic spin textures, \href{https://doi.org/10.1038/s41565-019-0383-4}{Nat. Nanotechnol. \textbf{14}, 328 (2019)}.
\bibitem{Park2021} H.-K. Park and S.-K. Kim, Channeling of spin waves in antiferromagnetic domain walls, \href{https://doi.org/10.1103/PhysRevB.103.214420}{Phys. Rev. B \textbf{103}, 214420 (2021)}.
\bibitem{Krawczyk2014} M. Krawczyk and D. Grundler, Review and prospects of magnonic crystals and devices with reprogrammable band structure, \href{https://doi.org/10.1088/0953-8984/26/12/123202}{J. Phys.: Condens. Matter \textbf{26}, 123202 (2014)}.
\bibitem{Chumak2017} A. V. Chumak, A. A. Serga, and B. Hillebrands, Magnonic crystals for data processing, \href{https://doi.org/10.1088/1361-6463/aa6a65}{J. Phys. D: Appl. Phys. \textbf{50}, 244001 (2017)}.
\bibitem{Zakeri2020} K. Zakeri, Magnonic crystals: towards terahertz frequencies, \href{https://doi.org/10.1088/1361-648X/ab88f2}{J. Phys.: Condens. Matter \textbf{32}, 363001 (2020)}.
\bibitem{Lv2025} Z. Lv, Z. Yan, Z. Li, X. Wang, Y. Nie, Q. Xia, X. Han, and G. Guo, Polarization-dependent spin wave channels in antiferromagnetic magnonic crystals, \href{https://doi.org/10.1063/5.0256440}{Appl. Phys. Lett. \textbf{126}, 122406 (2025)}.
\bibitem{Kim2009} S.-K. Kim, K.-S. Lee, and D.-S. Han, A gigahertz-range spin-wave filter composed of width-modulated nanostrip magnonic-crystal waveguides, \href{https://doi.org/10.1063/1.3186782}{Appl. Phys. Lett. \textbf{95}, 082507 (2009)}.
\bibitem{Merbouche2021} H. Merbouche, M. Collet, M. Evelt, V. E. Demidov, J. L. Prieto, M. Mu\~{n}oz, J. Ben Youssef, G. de Loubens, O. Klein, S. Xavier, O. D'Allivy Kelly, P. Bortolotti, V. Cros, A. Anane, and S. O. Demokritov, Frequency Filtering with a Magnonic Crystal Based on Nanometer-Thick Yttrium Iron Garnet Films, \href{https://doi.org/10.1021/acsanm.0c02382}{ACS Appl. Nano Mater. \textbf{4}, 121 (2021)}.
\bibitem{Wang2009} Z. K. Wang, V. L. Zhang, H. S. Lim, S. C. Ng, M. H. Kuok, S. Jain, and A. O. Adeyeye, Observation of frequency band gaps in a one-dimensional nanostructured magnonic crystal, \href{https://doi.org/10.1063/1.3089839}{Appl. Phys. Lett. \textbf{94}, 083112 (2009)}.
\bibitem{Ciubotaru2013} F. Ciubotaru, A. V. Chumak, B. Obry, A. A. Serga, and B. Hillebrands, Magnonic band gaps in waveguides with a periodic variation of the saturation magnetization, \href{https://doi.org/10.1103/PhysRevB.88.134406}{Phys. Rev. B \textbf{88}, 134406 (2013)}.
\bibitem{Li2016} Z. Li, M. Wang, Y. Nie, D. Wang, Q. Xia, W. Tang, Z. Zeng, and G. Guo, Spin-wave propagation spectrum in magnetization-modulated cylindrical nanowires, \href{https://doi.org/10.1016/j.jmmm.2016.04.057}{J. Magn. Magn. Mater. \textbf{414}, 49 (2016)}.
\bibitem{Qin2018} H. Qin, G.-J. Both, S. J. H\"{a}m\"{a}l\"{a}inen, L. Yao, and S. Van Dijken, Low-loss YIG-based magnonic crystals with large tunable bandgaps, \href{https://doi.org/10.1038/s41467-018-07893-5}{Nat. Commun. \textbf{9}, 5445 (2018)}.
\bibitem{Mieszczak2022} S. Mieszczak and J. W. K{\l}os, Interface modes in planar one-dimensional magnonic crystals, \href{https://doi.org/10.1038/s41598-022-15328-x}{Sci. Rep. \textbf{12}, 11335 (2022)}.
\bibitem{Adhikari2023} A. Adhikari, S. Majumder, Y. Otani, and A. Barman, Active Control of Dipole-Exchange Coupled Magnon Modes in Nanoscale Bicomponent Magnonic Crystals, \href{https://doi.org/10.1021/acsanm.2c05441}{ACS Appl. Nano Mater. \textbf{6}, 7166 (2023)}.
\bibitem{Lee2009} K.-S. Lee, D.-S. Han, and S.-K. Kim, Physical Origin and Generic Control of Magnonic Band Gaps of Dipole-Exchange Spin Waves in Width-Modulated Nanostrip Waveguides, \href{https://doi.org/10.1103/PhysRevLett.102.127202}{Phys. Rev. Lett. \textbf{102}, 127202 (2009)}.
\bibitem{Frey2020} P. Frey, A. A. Nikitin, D. A. Bozhko, S. A. Bunyaev, G. N. Kakazei, A. B. Ustinov, B. A. Kalinikos, F. Ciubotaru, A. V. Chumak, Q. Wang, V. S. Tiberkevich, B. Hillebrands, and A. A. Serga, Reflection-less width-modulated magnonic crystal, \href{https://doi.org/10.1038/s42005-020-0281-y}{Commun. Phys. \textbf{3}, 17 (2020)}.
\bibitem{Bir2024} A. S. Bir, S. V. Grishin, A. A. Grachev, O. I. Moskalenko, A. N. Pavlov, D. V. Romanenko, V. N. Skorokhodov, and S. A. Nikitov, Direct electric current control of hyperchaotic packets of dissipative dark envelope solitons in a magnonic crystal active ring resonator, \href{https://doi.org/10.1103/PhysRevApplied.21.044008}{Phys. Rev. Appl. \textbf{21}, 044008 (2024)}.
\bibitem{Mantion2024} S. Mantion, A. Torres Dias, M. Madami, S. Tacchi, and N. Biziere, Reconfigurable spin wave modes in a Heusler magnonic crystal, \href{https://doi.org/10.1063/5.0189486}{J. Appl. Phys. \textbf{135}, 053902 (2024)}.
\bibitem{Sadovnikov2022} A. V. Sadovnikov, G. Talmelli, G. Gubbiotti, E. N. Beginin, S. Sheshukova, S. A. Nikitov, C. Adelmann, and F. Ciubotaru, Reconfigurable 3D magnonic crystal: Tunable and localized spin-wave excitations in CoFeB meander-shaped film, \href{https://doi.org/10.1016/j.jmmm.2021.168670}{J. Magn. Magn. Mater. \textbf{544}, 168670 (2022)}.
\bibitem{Korniienko2019} A. Korniienko, V. P. Kravchuk, O. V. Pylypovskyi, D. D. Sheka, J. Van Den Brink, and Y. Gaididei, Curvature induced magnonic crystal in nanowires, \href{https://doi.org/10.21468/SciPostPhys.7.3.035}{SciPost Phys. \textbf{7}, 035 (2019)}.
\bibitem{Chumak2009} A. V. Chumak, T. Neumann, A. A. Serga, B. Hillebrands, and M. P. Kostylev, A current-controlled, dynamic magnonic crystal, \href{https://doi.org/10.1088/0022-3727/42/20/205005}{J. Phys. D: Appl. Phys. \textbf{42}, 205005 (2009)}.
\bibitem{Karenowska2012} A. D. Karenowska, J. F. Gregg, V. S. Tiberkevich, A. N. Slavin, A. V. Chumak, A. A. Serga, and B. Hillebrands, Oscillatory Energy Exchange between Waves Coupled by a Dynamic Artificial Crystal, \href{https://doi.org/10.1103/PhysRevLett.108.015505}{Phys. Rev. Lett. \textbf{108}, 015505 (2012)}.
\bibitem{Vogel2015} M. Vogel, A. V. Chumak, E. H. Waller, T. Langner, V. I. Vasyuchka, B. Hillebrands, and G. von Freymann, Optically reconfigurable magnetic materials, \href{https://doi.org/10.1038/nphys3325}{Nat. Phys. \textbf{11}, 487 (2015)}.
\bibitem{Wang2017} Q. Wang, A. V. Chumak, L. Jin, H. Zhang, B. Hillebrands, and Z. Zhong, Voltage-controlled nanoscale reconfigurable magnonic crystal, \href{https://doi.org/10.1103/PhysRevB.95.134433}{Phys. Rev. B \textbf{95}, 134433 (2017)}.
\bibitem{Chumak2010} A. V. Chumak, P. Dhagat, A. Jander, A. A. Serga, and B. Hillebrands, Reverse Doppler effect of magnons with negative group velocity scattered from a moving Bragg grating, \href{https://doi.org/10.1103/PhysRevB.81.140404}{Phys. Rev. B \textbf{81}, 140404(R) (2010)}.
\bibitem{Li2015} Z.-X. Li, X.-G. Wang, D.-W. Wang, Y.-Z. Nie, W. Tang, and G.-H. Guo, Reconfigurable magnonic crystal consisting of periodically distributed domain walls in a nanostrip, \href{https://doi.org/10.1016/j.jmmm.2015.04.012}{J. Magn. Magn. Mater. \textbf{388}, 10 (2015)}.
\bibitem{Wang2015} X.-G. Wang, G.-H. Guo, Z.-X. Li, D.-W. Wang, Y.-Z. Nie, and W. Tang, Spin-wave propagation in domain wall magnonic crystal, \href{https://doi.org/10.1209/0295-5075/109/37008}{EPL \textbf{109}, 37008 (2015)}.
\bibitem{Ma2015} F. Ma, Y. Zhou, H. B. Braun, and W. S. Lew, Skyrmion-Based Dynamic Magnonic Crystal, \href{https://doi.org/10.1021/acs.nanolett.5b00996}{Nano Lett. \textbf{15}, 4029 (2015)}.
\bibitem{Chen2021} Z. Chen and F. Ma, Skyrmion based magnonic crystals, \href{https://doi.org/10.1063/5.0061832}{ J. Appl. Phys. \textbf{130}, 090901 (2021)}.
\bibitem{Medlej2024} I. Medlej, J. Wang, C. Hu, and K. Yu, Hopfion based magnonic crystal, \href{https://doi.org/10.1016/j.jmmm.2024.171726}{J. Magn. Magn. Mater. \textbf{591}, 171726 (2024)}.
\bibitem{Wang2020} X.-G. Wang, Y.-Z. Nie, Q.-L. Xia, and G.-H. Guo, Dynamically reconfigurable magnonic crystal composed of artificial magnetic skyrmion lattice, \href{https://doi.org/10.1063/5.0012791}{J. Appl. Phys. \textbf{128}, 063901 (2020)}.
\bibitem{Li2018} Z.-X. Li, C. Wang, Yunshan Cao, and Peng Yan, Edge states in a two-dimensional honeycomb lattice of massive magnetic skyrmions, \href{https://doi.org/10.1103/PhysRevB.98.180407}{Phys. Rev. B \textbf{98}, 180407(R) (2018)}.
\bibitem{Li202102} Z.-X. Li, Z. Wang, Y. Cao, H. W. Zhang, and P. Yan, Robust edge states in magnetic soliton racetrack, \href{https://doi.org/10.1103/PhysRevB.103.054438}{Phys. Rev. B \textbf{103}, 054438 (2021)}.
\bibitem{Diaz2019} S. A. D\'{\i}az, J. Klinovaja, and D. Loss, Topological Magnons and Edge States in Antiferromagnetic Skyrmion Crystals, \href{https://doi.org/10.1103/PhysRevLett.122.187203}{Phys. Rev. Lett. \textbf{122}, 187203 (2019)}.
\bibitem{Hirosawa2020} T. Hirosawa, S. A. D\'{\i}az, J. Klinovaja, and D. Loss, Magnonic Quadrupole Topological Insulator in Antiskyrmion Crystals, \href{https://doi.org/10.1103/PhysRevLett.125.207204}{Phys. Rev. Lett. \textbf{125}, 207204 (2020)}.
\bibitem{Xie2021} K. Xie, L. Zhang, and F. Ma, Magnonic topological insulator realized in 2D magnetic skyrmion crystals, \href{https://doi.org/10.1063/5.0063972}{J. Appl. Phys. \textbf{130}, 153901 (2021)}.
\bibitem{Yi2024} Q. Yi, Z. Tang, D. Zhu, X. Xing, W. Zhang, and Y. Zhou, Magnonic bandgap openings and in-gap propagating states in domain-wall waveguides induced by periodic modulations, \href{https://doi.org/10.1103/PhysRevB.109.184401}{Phys. Rev. B \textbf{109}, 184401 (2024)}.
\bibitem{Malozemoff1972} A. P. Malozemoff, and J. C. Slonczewski, Effect of Bloch Lines on Magnetic Domain-Wall Mobility, \href{https://doi.org/10.1103/PhysRevLett.29.952}{Phys. Rev. Lett. \textbf{29}, 952 (1972)}.
\bibitem{Yoshimura2016} Y. Yoshimura, K.-J. Kim, T. Taniguchi, T. Tono, K. Ueda, R. Hiramatsu, T. Moriyama, K. Yamada, Y. Nakatani, and T. Ono, Soliton-like magnetic domain wall motion induced by the interfacial Dzyaloshinskii-Moriya interaction, \href{https://doi.org/10.1038/nphys3535}{Nat. Phys. \textbf{12}, 157 (2016)}.
\bibitem{Nagase2021} T. Nagase, Y. G. So, H. Yasui, T. Ishida, H. K. Yoshida, Y. Tanaka, K. Saitoh, N. Ikarashi, Y. Kawaguchi, M. Kuwahara, and M. Nagao, Observation of domain wall bimerons in chiral magnets, \href{https://doi.org/10.1038/s41467-021-23845-y}{Nat. Commun. \textbf{12}, 3490 (2021)}.
\bibitem{Chen2025} J. Chen, L. Shen, H. An, X. Zhang, H. Zhang, H. Du, X. Li, and Y. Zhou, Magnetic bimeron traveling on the domain wall, \href{https://doi.org/10.1063/5.0249927}{Appl. Phys. Lett. \textbf{126}, 142402 (2025)}.
\bibitem{Fu2025} W. L. Fu, H. M. Dong, and K. Chang, Tilted chiral spin textures in confined nanostructures with in-plane magnetic anisotropy, \href{https://doi.org/10.1103/PhysRevB.111.045422}{Phys. Rev. B \textbf{111}, 045422 (2025)}.
\bibitem{Cheng2019} R. Cheng, M. Li, A. Sapkota, A. Rai, A. Pokhrel, T. Mewes, C. Mewes, D. Xiao, M. De Graef, and V. Sokalski, Magnetic domain wall skyrmions, \href{https://doi.org/10.1103/PhysRevB.99.184412}{Phys. Rev. B \textbf{99}, 184412 (2019)}.
\bibitem{Li2021} M. Li, A. Rai, A. Pokhrel, A. Sapkota, C. Mewes, T. Mewes, D. Xiao, M. De Graef, and V. Sokalski, Magnetic domain wall substructures in Pt/Co/Ni/Ir multi-layers, \href{https://doi.org/10.1063/5.0056100}{J. Appl. Phys. \textbf{130}, 153903 (2021)}.
\bibitem{Yang2021} K. Yang, K. Nagase, Y. Hirayama, T. D. Mishima, M. B. Santos, and H. Liu, Wigner solids of domain wall skyrmions, \href{https://doi.org/10.1038/s41467-021-26306-8}{Nat. Commun. \textbf{12}, 6006 (2021)}.
\bibitem{Ross2023} C. Ross and M. Nitta, Domain-wall skyrmions in chiral magnets, \href{https://doi.org/10.1103/PhysRevB.107.024422}{Phys. Rev. B \textbf{107}, 024422 (2023)}.
\bibitem{Amari2024} Y. Amari, C. Ross, and M. Nitta, Domain-wall skyrmion chain and domain-wall bimerons in chiral magnets, \href{https://doi.org/10.1103/PhysRevB.109.104426}{Phys.  Rev. B \textbf{109}, 104426 (2024)}.
\bibitem{Han2024} S. U. Han, W. Kim, S. K. Kim, and S.-G. Je, Tunable domain-wall skyrmion Hall effect driven by a current and a magnetic field, \href{https://doi.org/10.1103/PhysRevB.109.014404}{Phys. Rev. B \textbf{109}, 014404 (2024)}.
\bibitem{Nie2025} H. Nie, Z. Li, X. Wang, and Z. Wang, Current-driven motion of magnetic domain wall skyrmions, \href{https://doi.org/10.1063/5.0250430}{Appl. Phys. Lett. \textbf{126}, 132402 (2025)}.
\bibitem{Xiao2025} A.-P. Xiao, L. Xiong, C. Yang, J.-Y. Jiang, and B. Zheng, Domain-wall skyrmion dynamics driven by current and waveguide spin wave: Synergistic acceleration effect, \href{https://doi.org/10.1103/PhysRevB.111.024402}{Phys. Rev. B \textbf{111}, 024402 (2025)}.
\bibitem{Saji2025} C. Saji, E. Saavedra, R. E. Troncoso, M. A. Castro, S. Allende, and A. S. Nunez, Magnonics along the wall in bimeron chain domain walls, \href{https://doi.org/10.1103/PhysRevB.111.174401}{Phys. Rev. B \textbf{111}, 174401 (2025)}.
\bibitem{Vansteenkiste2014} A. Vansteenkiste, J. Leliaert, M. Dvornik, M. Helsen, F. Garcia-Sanchez, and B. Van Waeyenberge, The design and verification of MuMax3, \href{https://doi.org/10.1063/1.4899186}{AIP Adv. \textbf{4}, 107133 (2014)}.
\bibitem{Ma2024} C. Ma, K.-J. Jin, E.-J. Guo, C. Ge, C. Wang, and X.-L. Xu, Dzyaloshinskii-Moriya interaction transistor with magnetization manipulated by electric field, \href{https://doi.org/10.1103/PhysRevB.110.094418}{Phys. Rev. B \textbf{110}, 094418 (2024)}.
\bibitem{Yu2023} D. Yu, Y. Ga, J. Liang, C. Jia, and H. Yang, Voltage-Controlled Dzyaloshinskii-Moriya Interaction Torque Switching of Perpendicular Magnetization, \href{https://doi.org/10.1103/PhysRevLett.130.056701}{Phys. Rev. Lett. \textbf{130}, 056701 (2023)}.
\bibitem{Yu2024} D. Yu, Y. Ga, P. Li, J. Jiang, J. Liang, L. Wang, C. Jia, K. Chang, and H. Yang, Voltage-Controlled Bimeron-Torques Switching of In-Plane Magnetization, \href{https://doi.org/10.1103/PhysRevLett.133.206701}{Phys. Rev. Lett. \textbf{133}, 206701 (2024)}.
\bibitem{Gallardo2019} R. A. Gallardo, D. Cort\'{e}s-Ortu\~{n}o, T. Schneider, A. Rold\'{a}n-Molina, F. Ma, R. E. Troncoso, K. Lenz, H. Fangohr, J. Lindner, and P. Landeros, Flat Bands, Indirect Gaps, and Unconventional Spin-Wave Behavior Induced by a Periodic Dzyaloshinskii-Moriya Interaction, \href{https://doi.org/10.1103/PhysRevLett.122.067204}{Phys. Rev. Lett. \textbf{122}, 067204 (2019)}.
\bibitem{Wei2025} F. Wei, Y. Zhou, W. Zhang, Z. Ren, G. Chen, H. Li, G. Han, S. Yan, and S. Kang, Flat band structure in chiral magnonic crystals with tunable indirect band gaps, \href{https://doi.org/10.1103/PhysRevApplied.23.024023}{Phys. Rev. Appl. \textbf{23}, 024023 (2025)}.
\bibitem{Holstein1940} T. Holstein and H. Primakoff, Field Dependence of the Intrinsic Domain Magnetization of a Ferromagnet, \href{https://doi.org/10.1103/PhysRev.58.1098}{Phys. Rev. \textbf{58}, 1098 (1940)}.
\bibitem{Colpa1978} J. H. P. Colpa, Diagonalization of the quadratic boson hamiltonian, \href{https://doi.org/10.1016/0378-4371(78)90160-7}{Physica A: Stat. Mech. Appl. \textbf{93}, 327 (1978)}.
\bibitem{Roldan2016} A. Rold\'{a}n-Molina, A. S. Nunez, and J. Fern\'{a}ndez-Rossier, Topological spin waves in the atomic-scale magnetic skyrmion crystal, \href{https://doi.org/10.1088/1367-2630/18/4/045015}{New J. Phys. \textbf{18}, 045015 (2016)}.
\bibitem{Takeda2024} H. Takeda, M. Kawano, K. Tamura, M. Akazawa, J. Yan, T. Waki, H. Nakamura, K. Sato, Y. Narumi, M. Hagiwara, M. Yamashita, and C. Hotta, Magnon thermal Hall effect via emergent SU(3) flux on the antiferromagnetic skyrmion lattice, \href{https://doi.org/10.1038/s41467-024-44793-3}{Nat. Commun. \textbf{15}, 566 (2024)}.
\end{thebibliography}
\end{document}